\documentclass[onecolumn,american]{IEEEtran}
\usepackage[T1]{fontenc}
\usepackage[utf8]{inputenc}
\usepackage{amsmath}
\usepackage{amsthm, hyperref}
\usepackage{amssymb}

\makeatletter
\theoremstyle{plain}
\newtheorem{thm}{\protect\theoremname}
\theoremstyle{definition}
\newtheorem{defn}{\protect\definitionname}
\theoremstyle{definition}
\newtheorem{remark}{\protect\remarkname}
\theoremstyle{plain}
\newtheorem{prop}{\protect\propositionname}
\theoremstyle{plain}
\newtheorem{lem}{\protect\lemmaname}
\theoremstyle{plain}
\newtheorem{cor}{\protect\corollaryname}

\usepackage[dvipsnames]{xcolor}

\makeatother

\usepackage{babel}
\providecommand{\corollaryname}{Corollary}
\providecommand{\definitionname}{Definition}
\providecommand{\remarkname}{Remark}
\providecommand{\lemmaname}{Lemma}
\providecommand{\propositionname}{Proposition}
\providecommand{\theoremname}{Theorem}

\begin{document}
\title{Lossless Source Coding with Marginal Distribution Constraint}
\author{%
Cheuk Ting Li${}^\S$, 
Jun Chen${}^\dagger$, and 
Shahab Asoodeh${}^\ddagger$
\\

\thanks{${}^\S$C. T. Li is with the Department of Information Engineering, The Chinese University of Hong Kong, Hong Kong, China (email: ctli@ie.cuhk.edu.hk).}
\thanks{${}^\dagger$J. Chen is with the Department of Electrical and Computer Engineering, McMaster University, Hamilton, ON L8S 4K1, Canada (email: {chenjun@mcmaster.ca}).}
\thanks{${}^\ddagger$S. Asoodeh is with Vector Institute and the Department of Computing and Software, McMaster University, Hamilton,
ON L8S 4K1, Canada (email: {asoodeh@mcmaster.ca}).}

}

\maketitle
\begin{abstract}
We consider lossless source coding with a marginal distribution constraint on the encoding symbols (SC-MDC). Unlike distribution matching, signal shaping and homophonic coding which require the symbols to be (approximately) i.i.d. following a target distribution, SC-MDC requires each symbol to have a marginal distribution that exactly matches the target distribution, which can be regarded as a perfect perceptual quality constraint on the symbols, and may have applications to wear leveling, source coding with cost and steganography.
We give nonasymptotic lower and upper bounds on the optimal encoding length, and exactly characterize the coding rate (or the first-order term of the length) when the source distribution or the target distribution is uniform, or when the source is an i.i.d. sequence.
\end{abstract}

\section{Introduction}

Consider a lossless source coding setting with a marginal distribution constraint (SC-MDC) on the encoding symbols. We encode a source $X\sim P$, supported on $\mathcal{X}$, into a sequence of encoding symbols $Y_{1},\ldots,Y_{m}$ from $\mathcal{Y}$ via a stochastic encoder $P_{Y_{1},\ldots,Y_{m}|X}$, so that the source can be recovered exactly, i.e., $H(X|Y_{1},\ldots,Y_{m})=0$. Unlike conventional source coding where we only impose a constraint on the alphabet $\mathcal{Y}$ (e.g., $\mathcal{Y}=\{0,1\}$), we impose a constraint on the marginal distribution that $Y_{i}\sim Q$ for $i\in[m]$, where $Q$ is a given target distribution. The goal is to find the smallest possible length $m$. 

This problem is equivalent to finding a coupling between $P$ and $m$ copies of $Q$ (i.e., $X\sim P$, $Y_{i}\sim Q$ for $i\in[m]$) satisfying $H(X|Y_{1},\ldots,Y_{m})=0$. The smallest $m$ is given by the following quantity which we call \emph{coupled aggregation number}:
\[
m^{*}(P\Vert Q):=\min\big\{ m\in\mathbb{Z}_{\ge0}:\,\exists\bar{Q}\in\Gamma_{m}(Q),\,\bar{Q}\sqsubseteq P\big\},
\]
where $\Gamma_{m}(Q)$ is the set of all couplings of $m$ copies of $Q$, i.e., the set of possible joint distributions of $Y_{1},\ldots,Y_{m}$ where $Y_{i}\sim Q$,\footnote{$\Gamma_{0}(Q)$ has only one element, which is the degenerate distribution at $\emptyset$ (the zero-length sequence).} and $\bar{Q}\sqsubseteq P$ ($P$ is an aggregation of $\bar{Q}$ \cite{vidyasagar2012metric,cicalese2016approximating,li2021efficient}) if there exists a function $f$ such that if $\bar{Y}\sim\bar{Q}$, then $f(\bar{Y})\sim P$.\footnote{We take $m^{*}(P\Vert Q)=\infty$ if no such $m$ exists.}

This quantity was introduced in the setting of learning with totally unpaired data \cite{li2026learning}, which concerns the smallest possible value of $H(X|Y_{1},\ldots,Y_{m})$ given the marginal distributions of $X,Y_1,\ldots,Y_m$, which corresponds to the smallest average cross entropy loss of predicting $X$ given the features $Y_1,\ldots,Y_m$ when only their marginals are known, i.e., when we only have $m+1$ unpaired datasets, one for each of $X,Y_1,\ldots,Y_m$. The value of $m^{*}(P\Vert Q)$ is the smallest number of identically distributed features $Y_1,\ldots,Y_m$ that allow perfect prediction of $X$, i.e., $H(X|Y_{1},\ldots,Y_{m})=0$.

This setting is also related to distribution matching \cite{kschischang1993optimal,bocherer2011matching,schulte2015constant}, signal shaping \cite{fischer2005precoding,mondelli2018achieve} and homophonic coding \cite{gunther1988universal,hoshi2001interval,Honda2016homophonic}, which require $Y_{1},\ldots,Y_{m}$ to be approximately i.i.d.\ following $Q$ (this is only satisfied weakly in the sense of vanishing normalized divergence for distribution matching \cite{bocherer2014informational,schulte2015constant}, or requires fixed-to-variable or variable-to-fixed codes for homophonic coding \cite{gunther1988universal,hoshi2001interval,Honda2016homophonic}; see Section \ref{sec:related} for details). 
We instead impose a marginal distribution constraint $Y_{i}\sim Q$, which must be satisfied exactly. We list some motivations for imposing marginal constraints:
\begin{itemize}
\item \emph{Encoding with perfect perceptual quality.} Suppose we encode $X\sim P$ into a sequence of images $Y_{1},\ldots,Y_{m}$, and we want each image to have a high perceptual quality. In the literature of rate-distortion-perception trade-off \cite{blau2018perception,blau2019rethinking,theis2021coding,chen2022rate}, $Y_{i}$ having perfect perceptual quality means $Y_{i}\sim Q$ where $Q$ is a natural distribution of images (this is referred to as \emph{weak-sense perfect perceptual quality} in \cite{chen2022rate}). More generally, if $Q$ is a natural distribution for the type of data $Y_{i}$ belongs to (e.g., audio, DNA strand, chemical solution), then $Y_{i}\sim Q$ would make $Y_{i}$ indistinguishable from a natural sample. We can also relax the perceptual constraint by only requiring the divergence  between $Q$ and the distribution of $Y_{i}$ to be small, and study the rate-perception trade-off; see Section \ref{sec:perception}. 
\item \emph{Wear leveling.} Consider a visible light communication setting where we use an array of $m$ LEDs ($Y_{i}$ is the intensity of the $i$-th LED) to transmit $X$ (e.g., see \cite{nagura2010improved}). Having the same marginal distribution ensures the same amount of lumen depreciation for each LED in the long run. If we apply on-off keying where $Y_{i}\in\{0,1\}$, we may want to ensure $Y_{i}\sim\mathrm{Bern}(\rho)$ for a small $\rho$, so the LEDs wear evenly.\footnote{Using a code with a small number of $1$'s (e.g., a constant composition code \cite{schulte2015constant}) is not sufficient if the same code is used repeatedly. For example, if the code sets $Y_{1}=1$ for most choices of $X$, then the first LED will wear quickly, even though the total number of $1$'s may be small.} This is also applicable to flash memory where we have $m$ cells, the wear of the $i$-th cell can depend on the level $Y_{i}$ programmed to the cell \cite{jagmohan2010adaptive}, and we want the cells to wear evenly. 
\item \emph{Unknown or adversarial cost per symbol.} Different values of $Y_{i}$ may require a different amount of energy or cost to store or transmit \cite{krause1962channels,gilbert1995coding}. Controlling the distribution of $Y_{i}$ allows us to limit the expected cost. SC-MDC is particularly useful when the cost function is unknown and may be different for each $Y_{i}$, or equivalently, the cost $c(Y_{i},z_{i})$ of $Y_{i}$ also depends on an unknown or adversarial state $z_{i}\in\mathcal{Z}$. For example, in flash memory \cite{jagmohan2010adaptive}, the cost (energy or wear) of writing to a cell depends not only on the target level $Y_{i}$, but also on its previous level $z_{i}$. We minimize the worst-case average cost 
\[
\bar{c}:=\max_{z^{m}\in\mathcal{Z}^{m}}\frac{1}{m}\mathbb{E}\bigg[\sum_{i=1}^{m}c(Y_{i},z_{i})\bigg].
\]
Since we are minimizing a worst-case expectation, a stochastic encoder (like a mixed strategy in game theory) is generally needed. If the minimax optimal distribution 
\begin{equation}
\underset{Q}{\mathrm{argmin}}\,\max_{z\in\mathcal{Z}}\mathbb{E}_{Y\sim Q}[c(Y,z)]\label{eq:minimax}
\end{equation}
is unique, then attaining the optimal worst-case average cost $\bar{c}$ is equivalent to fulfilling the marginal distribution constraint $Y_{i}\sim Q$ where $Q$ is the minimizer in (\ref{eq:minimax}) (assuming that the constraint can be fulfilled), since $\bar{c}=m^{-1}\sum_{i}\max_{z\in\mathcal{Z}}\mathbb{E}[c(Y_{i},z)]$ is minimized if and only if $Y_{i}\sim Q$.\footnote{The case where the minimizer is not unique is addressed in Section \ref{sec:perception}.}
For example, if $Y_{i},z_{i}\in\{0,1\}$ are the new and previous bits stored in a cell, respectively, and $c(1,0)=\alpha>0$ (it costs $\alpha$ to change $0$ to $1$), $c(0,1)=\beta>0$, $c(0,0)=c(1,1)=0$, then the unique optimal $Q$ is $\mathrm{Bern}(\beta/(\alpha+\beta))$, and we should ensure $Y_{i}\sim Q$. Other techniques such as constant composition codes \cite{schulte2015constant} would not give an exact optimality guarantee.  

\item \emph{Steganography with limited adversary.}
Consider a steganography system similar to \cite{cachin1998information,witt2023perfectly}, where an encoder wants to send a plaintext $T$ to a decoder by producing stegotexts $Y_1,\ldots,Y_m$. The encoder and decoder share a secret key $K$, and the plaintext must be recoverable from the stegotexts and the key, i.e., $H(T|Y_1,\ldots,Y_m,K)=0$. The goal is to ensure that a limited adversary who only observes any one stegotext $Y_i$ would not be able to know that communication is occurring. Assuming $Q$ is the covertext distribution (the distribution if no communication is occurring), we require $Y_i \sim Q$ for each $i$. One method is to encrypt $T$ into a uniformly-distributed ciphertext $C$ (e.g., via one-time pad \cite{shannon1949communication} $C=T \oplus K$ where $C,T,K \in \{0,1\}^{\ell}$, $K \sim \mathrm{Unif}(\{0,1\}^{\ell})$), and apply SC-MDC to encode $C$ into $Y_1,\ldots,Y_m$.\footnote{Steganography via minimum entropy coupling was studied in \cite{witt2023perfectly}. The difference is that we allow multiple stegotexts, where the adversary can only observe one of them. We also require $H(T|Y_1,\ldots,Y_m,K)=0$ exactly.}
This can also be regarded as a covert communication setting \cite{bloch2016covert}, where we ensure that a warden which only observes one $Y_i$ cannot detect the communication.

\end{itemize}
\smallskip{}

The number of encoding symbols needed by SC-MDC is $m^{*}(P\Vert Q)$, which can be interpreted as an operational measure of the ratio of the information in $P$ to the information in $Q$. One would expect
\begin{equation}
m^{*}(P\Vert Q)\approx\frac{H_{\alpha}(P)}{H_{\beta}(Q)},\label{eq:ratio}
\end{equation}
where $H_{\alpha},H_{\beta}$ are certain entropy measures. The question is whether they should be the Shannon entropy $H$, the Hartley entropy $H_{0}$ \cite{hartley1928transmission} or the min-entropy $H_{\infty}$ \cite{renyi1961measures}. We will prove various bounds on $m^{*}(P\Vert Q)$ that are indeed in the form of the ratio in (\ref{eq:ratio}). The choice of entropy measures would depend on the scenario, i.e., whether we consider arbitrary $P,Q$, or whether we fix $Q$ and let the entropy of $P$ become larger, or let $P$ be an i.i.d. product distribution with increasing length. To summarize the key results in this paper:
\begin{itemize}
\item For general $P,Q$ (Theorem \ref{thm:ratio}, Proposition \ref{prop:properties}),
\begin{equation}
\max\Big\{\frac{H(P)}{H(Q)},\frac{H_{0}(P)}{H_{0}(Q)}\Big\}\le m^{*}(P\Vert Q)<\frac{H_{0}(P)+1}{H_{\infty}(Q)}+3.\label{eq:bound_nonasymp}
\end{equation}
The upper bound is approximately $H_{0}(P)/H_{\infty}(Q)$, which is the ``most conservative'' bound in the form $H_{\alpha}(P)/H_{\beta}(Q)$ since $H_{0}$ is the largest and $H_{\infty}$ is the smallest among $H,H_{0},H_{\infty}$. The upper and lower bounds do not match unless $Q$ is uniform. For the special case $Q=\mathrm{Bern}(1/2)$, the bounds approximately match, and we can show that
\[
\left\lceil H_{0}(P)\right\rceil \le m^{*}(P\Vert\mathrm{Bern}(1/2))\le\left\lceil H_{0}(P)\right\rceil +1.
\]
Hence, the number of bits needed to encode $X\sim P$, if we require each bit to be $\mathrm{Bern}(1/2)$, is approximately $H_{0}(P)$ within $2$ bits. Comparing this to the classical result \cite{hartley1928transmission} that the number of bits needed without the marginal constraint is approximately $H_{0}(P)$ within $1$ bit, we can see that the marginal constraint only adds a penalty of $1$ bit. 
\item If $H_{0}(P)$ is large, we have a better upper bound (Theorem \ref{thm:H0_over_H})
\begin{equation}
m^{*}(P\Vert Q)\le\frac{H_{0}(P)}{H(Q)}+O\big((H_{0}(P))^{2/3}\big),\label{eq:bound_large_H0P}
\end{equation}
where the constant in $O(\cdot)$ only depends on $Q$. This upper bound $\approx H_{0}(P)/H(Q)$ is better than the $\approx H_{0}(P)/H_{\infty}(Q)$ in (\ref{eq:bound_nonasymp}), at the expense of having an $O(\cdot)$ term that makes the bound usable only when $H_{0}(P)\gg H(Q)$, which is a reasonable assumption for source coding since the whole source sequence usually has a much larger entropy than one encoding symbol. The upper bound (\ref{eq:bound_large_H0P}) approximately matches the lower bound in (\ref{eq:bound_nonasymp}) when $P$ is uniform, which yields
\[
m^{*}(\mathrm{Bern}(1/2)^{\otimes n}\Vert Q)=\frac{n}{H(Q)}+O(n^{2/3}).
\]
This gives a new operational meaning of entropy, as $H(Q)=\lim_{n\to\infty}n/m^{*}(\mathrm{Bern}(1/2)^{\otimes n}\Vert Q)$ is  the asymptotic number of i.i.d. source bits that can be encoded per symbol $Y_{i}\sim Q$.
\item For a source distribution $P^{\otimes n}$ which is an i.i.d. product distribution, we have (Theorem \ref{thm:gamma})
\begin{equation}
m^{*}(P^{\otimes n}\Vert Q)=n\max\Big\{\frac{H(P)}{H(Q)},\frac{H_{0}(P)}{H_{0}(Q)}\Big\}+o(n).\label{eq:intro_rate}
\end{equation}
This shows the asymptotic tightness of the lower bound in (\ref{eq:bound_nonasymp}), and is asymptotically tighter than (\ref{eq:bound_large_H0P}) at the expense of requiring an i.i.d. product distribution. In particular, for $P$ with finite support, we have
\[
m^{*}(P^{\otimes n}\Vert\mathrm{Geom}(1/2))=\frac{nH(P)}{2}+o(n).
\]
This gives another new operational meaning of entropy via the following source coding setting. Suppose we encode $X^{n}\sim P^{\otimes n}$ into $Y_{1},\ldots,Y_{\lfloor nR\rfloor}\in\mathbb{Z}_{\ge0}$, where we have a cost constraint $\mathbb{E}[Y_{i}]\le1$ for every $i$. The optimal rate $R$ as $n\to\infty$ is $H(P)/2$. Interestingly, the optimal rate is given in terms of entropy, even though the setting is fixed-length and zero-error.
\item Continuing the i.i.d. setting for general $Q$, if we allow a vanishing error probability $\epsilon$ in the source coding setting, or equivalently, allow a vanishing total variation distance from $P^{\otimes n}$, then the optimal coding rate is (Theorem \ref{thm:vanishing_error})
\begin{equation}
\lim_{\epsilon\to0}\underset{n\to\infty}{\mathrm{limsup}}\frac{1}{n}\inf_{\hat{P}:d_{\mathrm{TV}}(P^{\otimes n},\hat{P})\le\epsilon}m^{*}(\hat{P}\Vert Q)=\frac{H(P)}{H(Q)}.\label{eq:asymp}
\end{equation}
\item Continuing the i.i.d. setting, if we instead allow $Y_{i}$ to only approximately follow $Q$ by imposing a constraint on the divergence $\delta(Q,P_{Y_{i}})\le\Delta$ (which can be regarded as a perception constraint) where $\delta$ is convex in its second argument, then the optimal coding rate is (Theorem \ref{thm:perception})
\[
\max\bigg\{\frac{H(P)}{\sup_{\hat{Q}:\delta(Q,\hat{Q})\le\Delta}H(\hat{Q})},\,\frac{H_{0}(P)}{\sup_{\hat{Q}:\delta(Q,\hat{Q})\le\Delta}H_{0}(\hat{Q})}\bigg\},
\]
which generalizes (\ref{eq:intro_rate}). We call this the \emph{SC-MDC rate-perception function}.
\end{itemize}
\smallskip{}

\subsection*{Notation}

Logarithms are to the base $2$. Entropy is in bits. Write $\mathbb{N}:=\{1,2,\ldots\}$, $[a:b]:=\{a,a+1,\ldots,b\}$, $[n]:=[1:n]$. For a sequence $X_{1},\ldots,X_{n}$, write $X^{n}:=(X_{1},\ldots,X_{n})$. For discrete distributions $P,Q$, we say that $Q$ is \emph{majorized} by $P$ \cite{marshall1979inequalities}, denoted as $Q\preceq P$, if $\sum_{i=1}^{k}Q^{\downarrow}(i)\le\sum_{i=1}^{k}P^{\downarrow}(i)$ for every $k\in\mathbb{N}$, where $P^{\downarrow}(i)$ is the $i$-th largest entry of $P$ ($P^{\downarrow}(i)=0$ if $i$ is larger than the cardinality of $P$). For a discrete random variable $X$, write $P_{X}$ for its probability mass function. If $X$ is a random variable and $P_{Y|X}$ is a conditional distribution, ``generate $Y|X\sim P_{Y|X}$'' means we generate a random variable $Y$ conditional on $X$ according to $P_{Y|X}$, conditionally independently of all previously defined random variables given $X$. The uniform distribution over $\mathcal{A}$ is denoted as $\mathrm{Unif}(\mathcal{A})$ (write $\mathrm{Unif}[n]:=\mathrm{Unif}([n])$). The Bernoulli and geometric distributions are denoted as $\mathrm{Bern}(\alpha)$ and $\mathrm{Geom}(\alpha)$, respectively. The Hartley entropy \cite{hartley1928transmission} is $H_{0}(P):=\log|\{x:P(x)>0\}|$, and the min-entropy \cite{renyi1961measures} is $H_{\infty}(P):=-\log\max_{x}P(x)$.

\section{Related Works}\label{sec:related}

A relevant prior work is learning with totally unpaired data \cite{li2026learning}, where the following \emph{minimum conditional entropy coupling} problem is studied: minimize $H(X|Y_1,\ldots,Y_m)$ subject to the marginal distribution constraints $X \sim P$ and $Y_i \sim Q_i$ for $i=1,\ldots,m$. The optimal value is characterized when $P,Q_i$ are Bernoulli distributions. The settings and applications of \cite{li2026learning} and this paper are different. \cite{li2026learning} studies learning with totally unpaired data, and hence mostly focuses on the heterogeneous setting where $Y_i \sim Q_i$ may have different marginal distributions since each $Y_i$ is a different feature, and does not require $H(X|Y^m)$ to be $0$. In comparison, this paper focuses on source coding with marginal constraint, and hence studies the homogeneous setting where the $Y_i$'s have a common marginal distribution $Q$ since the encoding symbols are usually homogeneous in source coding settings, and requires the source to be decoded (i.e., $H(X|Y^m)=0$). Some results in \cite{li2026learning} apply to the homogeneous setting (e.g., \cite[Theorems 4, 5]{li2026learning}), though these results are not strong enough to yield the source coding results in \eqref{eq:bound_nonasymp},  \eqref{eq:bound_large_H0P}, \eqref{eq:intro_rate}.\footnote{\cite[Theorem 4]{li2026learning} shows an exponentially decreasing bound on the minimum conditional entropy, which is not sufficient for the source coding setting where the conditional entropy must be $0$ (see Remark \ref{rem:prev_exponential}). \cite[Theorem 5]{li2026learning} shows the existence of $m$ with zero conditional entropy, though it does not provide a bound on $m$.}
The setting of source coding with marginal distribution constraint gives different operational limits compared to \cite{li2026learning}, and requires completely different analysis.

Distribution matching \cite{kschischang1993optimal,bocherer2011matching,schulte2015constant,bocherer2014informational} concerns the problem of mapping an i.i.d. source $X_{1},\ldots,X_{n}\stackrel{\mathrm{iid}}{\sim}P$ ($P$ is usually taken to be $\mathrm{Bern}(1/2)$) to a sequence $(\tilde{Y}_{1},\ldots,\tilde{Y}_{m})=f(X_{1},\ldots,X_{n})$ (where $f$ is injective) that is approximately i.i.d. following a desired distribution $Q$. It has been shown in \cite{schulte2015constant} that for distribution matching with $P=\mathrm{Bern}(1/2)$, $m=\lfloor nR\rfloor$ and a vanishing normalized divergence constraint $n^{-1}D(P_{\tilde{Y}^{m}}\Vert Q^{\otimes m})\to0$, the optimal rate is $R^{*}=H(P)/H(Q)$. 
Homophonic codes can ensure i.i.d. output that follows $Q$ exactly, but only via fixed-to-variable codes \cite{hoshi2001interval} ($m$ depends on $X^n$) or variable-to-fixed codes \cite{Honda2016homophonic} ($n$ depends on $X_1,X_2,\ldots$), not fixed-to-fixed codes ($n,m$ are fixed).

In comparison, the SC-MDC setting only requires $Y_{1},\ldots,Y_{m}$ to be identically distributed (not i.i.d.) according to $Q$. Although the lower bound in (\ref{eq:bound_nonasymp}) shows that SC-MDC cannot achieve a rate better than the rate $H(P)/H(Q)$ of distribution matching, the operational advantage of SC-MDC is twofold. First, the distribution constraint in SC-MDC is exact, whereas the constraint in distribution matching is approximate. Distribution matching only guarantees a vanishing normalized divergence (e.g., \cite{schulte2015constant} gives $D(P_{\tilde{Y}^{m}}\Vert Q^{\otimes m})=O(\log n)$), which does not imply a vanishing total variation distance between $Q$ and the distribution of $\tilde{Y}_{i}$ (which can even be close to $1$). Second, SC-MDC allows one-shot achievability bounds like (\ref{eq:bound_nonasymp}) which apply to sources with small entropy, whereas theoretical results in distribution matching are often asymptotic or require a large $n$ (e.g., \cite{schulte2015constant}). 

These differences make the applications of distribution matching and SC-MDC vastly different. Distribution matching is useful for channel coding where we want the input sequence to be weakly approximately i.i.d. following the optimal distribution $Q$, but it is not important whether each individual symbol is exactly (or even approximately) following $Q$, since wasting $o(n)$ symbols would not affect the rate. SC-MDC is useful for source coding settings where it is important for each encoding symbol to follow $Q$ exactly, but the symbols being i.i.d. is not required. 

The infinite divisibility of information \cite{li2022infinite} concerns a setting where a source $X$ is encoded into an i.i.d. sequence $Y_{1},\ldots,Y_{m}$ with $H(X|Y_{1},\ldots,Y_{m})=0$. While the theoretical setting is similar to SC-MDC, the key difference is that \cite{li2022infinite} allows the encoder to choose the alphabet and distribution of $Y_{i}$ as a part of the scheme (the alphabet may be large), whereas the distribution $Q$ is given in SC-MDC, but we do not require $Y_{i}$'s to be i.i.d. Hence, \cite{li2022infinite} is not applicable to source coding settings where the alphabet of $Y_{i}$ is fixed.

Distribution-preserving and output-constrained quantization \cite{li2010distribution,saldi2014randomized} and rate-distortion-perception coding \cite{blau2018perception,blau2019rethinking,theis2021coding,chen2022rate} concern lossy source coding where the distribution of the reconstruction is controlled. This is different from SC-MDC, which controls the marginal distribution of the encoding symbols $Y_{i}$, not the reconstruction (which is the source $X$ since we consider lossless compression). As mentioned in the introduction, SC-MDC can be regarded as lossless source coding with perfect perceptual quality of the encoding symbols (not the reconstruction). 

The joint mixability problem \cite{wang2016joint} (also see \cite{bernard2018rearrangement}) studies whether there exists a coupling $(Y_{1},\ldots,Y_{n})$ of $Q_{1},\ldots,Q_{n}$ (i.e., $Y_{i}\sim Q_{i}$) such that $\sum_{i}Y_{i}\sim P$ for given $P,Q_{1},\ldots,Q_{n}$. In comparison, SC-MDC allows an arbitrary decoding function $X=g(Y_{1},\ldots,Y_{m})\sim P$, not only the sum. 

\smallskip{}

\section{General Bounds}

As mentioned in the introduction, the answer to the source coding with marginal distribution constraint (SC-MDC) problem is given by the coupled aggregation ratio defined as follows.

\smallskip{}

\begin{defn}
Given discrete distributions $P,Q_{1},\ldots,Q_{m}$, we say that $P$ is a \emph{coupled aggregation} of $Q_{1},\ldots,Q_{m}$, denoted as $(Q_{1},\ldots,Q_{m})\sqsubseteq P$, if there exists a coupling $\bar{Q}\in\Gamma(Q_{1},\ldots,Q_{m})$ such that $\bar{Q}\sqsubseteq P$, or equivalently, there exist random variables $X,Y_{1},\ldots,Y_{m}$ with marginal distributions $X\sim P$ and $Y_{i}\sim Q_{i}$ for $i\in[m]$ such that $H(X|Y_{1},\ldots,Y_{m})=0$. The \emph{coupled aggregation ratio} between $P$ and $Q$ is
\[
m^{*}(P\Vert Q):=\min\big\{ m\in\mathbb{Z}_{\ge0}:\,(\underbrace{Q,\ldots,Q}_{m\;\text{copies}})\sqsubseteq P\big\}.
\]
\end{defn}
\smallskip{}

We list some simple properties of the coupled aggregation ratio. The proof is in Appendix \ref{subsec:pf_properties}.

\smallskip{}

\begin{prop}
\label{prop:properties}For discrete distributions $P,Q,R$ and discrete random variables $X,Y,Z$,
\begin{itemize}
\item (Lower bounds).
\begin{equation}
m^{*}(P\Vert Q)\ge\max\Big\{\Big\lceil\frac{H(P)}{H(Q)}\Big\rceil,\Big\lceil\frac{H_{0}(P)}{H_{0}(Q)}\Big\rceil\Big\}.\label{eq:prop_lower_bound}
\end{equation}
Also,
\begin{equation}
m^{*}(P\Vert Q)\ge\Big\lceil\frac{1-\max_{x}P(x)}{1-\max_{x}Q(x)}\Big\rceil=\Big\lceil\frac{1-2^{-H_{\infty}(P)}}{1-2^{-H_{\infty}(Q)}}\Big\rceil.\label{eq:prop_lower_bound_infty}
\end{equation}
\item (Chain rule). $m^{*}(P\Vert Q)\le m^{*}(P\Vert R)\cdot m^{*}(R\Vert Q)$.
\item (Subadditivity). $m^{*}(P_{X,Z}\Vert Q)\le m^{*}(P_{X}\Vert Q)+m^{*}(P_{Z}\Vert Q)$.
\item (Almost quasiconvexity). For discrete random variables $X,Y,Z$,
\[
m^{*}(P_{X}\Vert P_{Y})\le\max_{z\in\mathcal{Z}}m^{*}(P_{X|Z=z}\Vert P_{Y|Z=z})+m^{*}(P_{Z}\Vert P_{Y}).
\]
\end{itemize}
\end{prop}
\smallskip{}

We can regard $m^{*}(P\Vert Q)$ as the ``ratio of the information in $P$ to the information in $Q$''. The lower bound (\ref{eq:prop_lower_bound}) shows that $m^{*}(P\Vert Q)\ge H_{\alpha}(P)/H_{\alpha}(Q)$ for $H_{\alpha}$ being $H$ or $H_{0}$. The lower bound (\ref{eq:prop_lower_bound_infty}) shows that $m^{*}(P\Vert Q)$ is approximately lower-bounded by $H_{\infty}(P)/H_{\infty}(Q)$ as well when $H_{\infty}(P),H_{\infty}(Q)$ are small.

Most of this paper is devoted to proving upper bounds on $m^{*}(P\Vert Q)$, which correspond to achievability results for SC-MDC. 
We first show a simple bound when $Q$ is a uniform distribution.

\smallskip{}

\begin{prop}\label{prop:Q_unif}
If $Q$ is a uniform distribution over a finite set,
\[
\left\lceil \frac{H_{0}(P)}{H_0(Q)}\right\rceil \le m^{*}(P\Vert Q)\le\left\lceil \frac{H_{0}(P)}{H_0(Q)}\right\rceil +1.
\]
\end{prop}
\begin{IEEEproof}
The lower bound follows from Proposition \ref{prop:properties}. For the upper bound, assume $Q = \mathrm{Unif}([0:b-1])$, $X \sim P$, and let $m = \lceil H_0(P)/H_0(Q) \rceil + 1$. Since $|\mathrm{supp}(P)| = 2^{H_0(P)} \le 2^{(m-1)H_0(Q)} = b^{m-1}$, there exists an injective mapping from $X$ to $Z_1,\ldots,Z_{m-1} \in [0:b-1]$, i.e., $H(X|Z^{m-1})=0$. Let $U \sim \mathrm{Unif}([0:b-1])$, and take $(Y_1,\ldots,Y_m) = (Z_1 \oplus U ,\ldots,Z_{m-1} \oplus U, U)$, where $\oplus$ denote addition modulo $b$. We can check that $Y_i \sim Q$, and $H(X|Y^m)=0$.
\end{IEEEproof}
\smallskip{}

To prove bounds for general $Q$, we will utilize the majorized alias lemma in \cite[Lemma 1]{li2021efficient} (also see \cite{cicalese2019minimum} for a more general construction). We state the lemma in the following convenient form. The proof that \cite[Lemma 1]{li2021efficient} implies Lemma \ref{lem:alias} is in Appendix \ref{subsec:pf_alias}.

\smallskip{}

\begin{lem}
[Majorized alias lemma \cite{li2021efficient,cicalese2019minimum}]\label{lem:alias}For discrete distributions $R\preceq P$ where $P$ is finite, there exists $0\le\rho\le1/2$ such that $(R,\mathrm{Bern}(\rho))\sqsubseteq P$.
\end{lem}
\smallskip{}

We present a key technical lemma about coupled aggregation, which utilizes the majorized alias lemma.

\smallskip{}

\begin{lem}
\label{lem:bern_coupling}Let $0\le a\le b\le1/2$ with $2a\ge b$ and $a+b\ge1/2$. For every $0\le\rho\le1$, we have $(\mathrm{Bern}(a),\mathrm{Bern}(b))\sqsubseteq\mathrm{Bern}(\rho)$. Moreover, for discrete distributions $R\preceq P$ where $P$ is finite, we have $(R,\mathrm{Bern}(a),\mathrm{Bern}(b))\sqsubseteq P$.
\end{lem}
\smallskip{}

\begin{IEEEproof}
Assume $0\le\rho\le1/2$. Consider a coupling $X\sim\mathrm{Bern}(a)$, $Y\sim\mathrm{Bern}(b)$ with joint probability mass function $p(1,1)=t$, $p(0,1)=b-t$, $p(1,0)=a-t$, $p(0,0)=1-a-b+t$ where $t\in[0,a]$. By varying $t$, we can make $p(1,1)$ any number in $[0,a]$, or make $p(0,1)$ any number in $[b-a,b]$, or make $1-p(0,0)$ any number in $[b,a+b]$. The union of these three intervals covers $[0,1/2]$ as long as $2a\ge b$ and $a+b\ge1/2$, so $\rho$ must fall in one of these intervals, and we can have a function of $(X,Y)$ which follows $\mathrm{Bern}(\rho)$. For the second claim, Lemma \ref{lem:alias} gives $(R,\mathrm{Bern}(\rho))\sqsubseteq P$ for some $\rho$, so the result follows from $(\mathrm{Bern}(a),\mathrm{Bern}(b))\sqsubseteq\mathrm{Bern}(\rho)$.
\end{IEEEproof}
\smallskip{}

Using Lemma \ref{lem:bern_coupling}, we can give the following upper bound on $m^{*}(P\Vert Q)$.

\smallskip{}

\begin{lem}
\label{lem:k_plus_2}If  there exists a coupling $R\in\Gamma_{k}(Q)$ with $R\preceq P$, then 
\[
m^{*}(P\Vert Q)\le k+\max\Big\{\Big\lceil\frac{1/2}{1-\max_{x}Q(x)}\Big\rceil,\,2\Big\}.
\]
\end{lem}
\smallskip{}

\begin{IEEEproof}
Without loss of generality, assume $Q$ is supported over $\mathbb{N}$ and is non-increasing. First consider the case $\rho:=1-\max_{x}Q(x)\ge1/4$. If $1/4\le\rho\le1/2$, since $Q\sqsubseteq\mathrm{Bern}(\rho)$, applying Lemma \ref{lem:bern_coupling} on $a=b=\rho$ gives $(R,\mathrm{Bern}(\rho),\mathrm{Bern}(\rho))\sqsubseteq P$, and $m^{*}(P\Vert Q)\le k+2$. If $1/2<\rho\le3/4$, we apply Lemma \ref{lem:bern_coupling} on $a=b=1-\rho\in[1/4,1/2]$ instead. If $\rho>3/4$, let $n^{*}:=\min\{n:\sum_{x=1}^{n}Q(x)\ge1/4\}$. Since $\max_{x}Q(x)=1-\rho<1/4$, we have $\sum_{x=1}^{n^{*}}Q(x)\in[1/4,1/2]$, so we can apply Lemma \ref{lem:bern_coupling} on $a=b=\sum_{x=1}^{n^{*}}Q(x)$.

We then consider the case $\rho<1/4$. We have $Q\sqsubseteq\mathrm{Bern}(\rho)$. Note that $(\mathrm{Bern}(\rho),\ldots,\mathrm{Bern}(\rho))\sqsubseteq\mathrm{Bern}(n\rho)$ as long as $n\rho\le1$, where there are $n$ copies in $(\mathrm{Bern}(\rho),\ldots,\mathrm{Bern}(\rho))$. This is because we can couple $Y_{1},\ldots,Y_{n}\sim\mathrm{Bern}(\rho)$ such that at most one $Y_{i}$ is $1$ at a time, and the probability that there exists $Y_{i}=1$ is $n\rho$. Let $\ell:=\lceil1/(2\rho)\rceil$. Applying Lemma \ref{lem:bern_coupling} on $a=\lfloor\ell/2\rfloor\rho$, $b=\lceil\ell/2\rceil\rho$ (we have $2a\ge\rho(\ell-1)/2+\rho\ge b$ and $a+b=\ell\rho\ge1/2$), we have $(R,\mathrm{Bern}(\lfloor\ell/2\rfloor\rho),\mathrm{Bern}(\lceil\ell/2\rceil\rho))\sqsubseteq P$, and hence $m^{*}(P\Vert Q)\le k+\lfloor\ell/2\rfloor+\lceil\ell/2\rceil=k+\ell$. 
\end{IEEEproof}
\smallskip{}

The following simple bound shows that $m^{*}(P\Vert Q)$ is approximately upper-bounded by $H_{0}(P)/H_{\infty}(Q)$. Since $H_{0}(P)$ is an ``over-estimate'' of the information in $P$, and $H_{\infty}(Q)$ is an ``under-estimate'' of the information in $Q$, the ratio $H_{0}(P)/H_{\infty}(Q)$ is a conservative upper bound on the ratio between the information in $P$ and that in $Q$.

\smallskip{}

\begin{thm}
\label{thm:ratio}We have
\begin{align*}
m^{*}(P\Vert Q) & \le\Big\lceil\frac{H_{0}(P)}{H_{\infty}(Q)}\Big\rceil+\max\Big\{\frac{1}{H_{\infty}(Q)},\,1\Big\}+1\\
 & <\frac{H_{0}(P)+1}{H_{\infty}(Q)}+3.
\end{align*}
Furthermore, we can find a coupling achieving the bound, where all except at most $1/H_{\infty}(Q)+2$ symbols in $Y_{1},\ldots,Y_{m}$ are i.i.d.
\end{thm}
\smallskip{}

\begin{IEEEproof}
Let $k:=\lceil H_{0}(P)/H_{\infty}(Q)\rceil$. We have $\max_{x^{k}}Q^{\otimes k}(x^{k})=2^{-kH_{\infty}(Q)}\le2^{-H_{0}(P)}=1/|\mathrm{supp}(P)|$, where $\mathrm{supp}(P)$ is the support of $P$. Hence, $Q^{\otimes k}\preceq\mathrm{Unif}(\mathrm{supp}(P))\preceq P$. If $H_{\infty}(Q)\le1$, we have $1-\max_{x}Q(x)=1-2^{-H_{\infty}(Q)}\ge H_{\infty}(Q)/2$. If $H_{\infty}(Q)>1$, then $1-\max_{x}Q(x)>1/2$. In both cases, we can show by Lemma \ref{lem:k_plus_2} that
\begin{align*}
m^{*}(P\Vert Q) & \le k+\max\Big\{\Big\lceil\frac{1/2}{1-\max_{x}Q(x)}\Big\rceil,\,2\Big\}\\
 & \le k+\max\Big\{\frac{1}{H_{\infty}(Q)}+1,\,2\Big\}\\
 & <\frac{H_{0}(P)+1}{H_{\infty}(Q)}+3.
\end{align*}
\end{IEEEproof}
\smallskip{}




We can also consider the heterogeneous case where we require $Y_{i}\sim Q_{i}$, and $Q_{1},\ldots,Q_{m}$ may be different.

\smallskip{}

\begin{thm}\label{thm:heterogeneous}
Given discrete distributions $P,Q_{1},\ldots,Q_{m}$, we have $(Q_{1},\ldots,Q_{m})\sqsubseteq P$ if there exists a partition $A_{0},A_{1},A_{2}$ of $[m]$ with $\sum_{i\in A_{j}}H_{\infty}(Q_{i})\ge\log(4/3)$ for $j\in\{1,2\}$, and $\sum_{i\in A_{0}}H_{\infty}(Q_{i})\ge H_{0}(P)$. For a simpler sufficient condition, we have $(Q_{1},\ldots,Q_{m})\sqsubseteq P$ if $H_{\infty}(Q_{i})$ is in ascending order, and
\[
\sum_{i=1}^{m-2}H_{\infty}(Q_{i})\ge H_{0}(P)+2\log\frac{4}{3}.
\]
\end{thm}
\smallskip{}

\begin{IEEEproof}
Consider $j\in\{1,2\}$, $\bar{Q}_{j}:=\bigotimes_{i\in A_{j}}Q_{i}$, with $H_{\infty}(\bar{Q}_{j})\ge\log(4/3)$, $\max_{x}\bar{Q}_{j}(x)\le3/4$. By the same arguments as in the proof of Lemma \ref{lem:k_plus_2}, we have $\bar{Q}_{j}\sqsubseteq\mathrm{Bern}(\lambda_{j})$ for some $\lambda_{j}\in[1/4,1/2]$. Since $\sum_{i\in A_{0}}H_{\infty}(Q_{i})\ge H_{0}(P)$, we have $\bigotimes_{i\in A_{0}}Q_{i}\preceq P$ (see the proof of Theorem \ref{thm:ratio}). Applying Lemma \ref{lem:bern_coupling} on $\bigotimes_{i\in A_{0}}Q_{i}$ and $a,b$ being $\lambda_{1},\lambda_{2}$, we have $(Q_{1},\ldots,Q_{m})\sqsubseteq P$.

For the simpler condition, let $t_{i}:=H_{\infty}(Q_{i})$, and assume $t_{i}$ is in ascending order, and $\sum_{i=1}^{m}t_{i}\ge H_{0}(P)+2\log(4/3)+t_{m}+t_{m-1}$. Assign $i=1,\ldots,m$ to $A_{0},A_{1},A_{2}$ by first filling up $A_{1}$ until $\sum_{i\in A_{1}}t_{i}\ge\log(4/3)$, and then filling up $A_{2}$ until $\sum_{i\in A_{2}}t_{i}\ge\log(4/3)$ (which is guaranteed to succeed due to the assumption), and then assign the remaining $i$'s to $A_{0}$. We have $\sum_{i\in A_{1}\cup A_{2}}t_{i}\le2\log(4/3)+t_{m}+t_{m-1}$, so $\sum_{i\in A_{0}}t_{i}\ge H_{0}(P)$.
\end{IEEEproof}
\smallskip{}

\begin{remark}\label{rem:prev_exponential}
In \cite[Theorem 4]{li2026learning}, letting $H^* = H^*(P \Vert Q_1,\ldots,Q_m) := \inf H(X|Y_1,\ldots,Y_m)$ where the infimum is taken over jointly-distributed random variables $(X,Y_1,\ldots,Y_m)$ with marginals $P,Q_1,\ldots,Q_m$, it was proved that 
\begin{align}
H^* & \le \min \big\{1, (2^{H_0(P)}-1) 2^{-\sum_{i=1}^m H_{\infty}(Q_i)}  \big\} \cdot H_0(P). \label{eq:prev_exponential}
\end{align}
This bound shows that $H^*$ is small when $\sum_{i=1}^m H_{\infty}(Q_i) - H_0(P)$ is large, which is order-wise similar to Theorem \ref{thm:ratio} (for the homogeneous case) and Theorem \ref{thm:heterogeneous}.
Nevertheless, the bound \eqref{eq:prev_exponential} only shows that $H^*$ decays exponentially as $m$ increases. Unlike Theorems \ref{thm:ratio} and \ref{thm:heterogeneous}, \eqref{eq:prev_exponential} cannot show that $H^*=0$ for $m$ large enough. 

\end{remark}

\section{Bounds for $P$ with Large Entropy}

Can the $H_{0}(P)/H_{\infty}(Q)$ upper bound in Theorem \ref{thm:ratio} be improved? For simplicity, we want the bound to be in the form $H_{\alpha}(P)/H_{\beta}(Q)$ for some entropy measures $H_{\alpha},H_{\beta}$. By Proposition \ref{prop:Q_unif}, we know that $H_{0}(P)$ is the correct entropy measure for $P$ when $Q=\mathrm{Bern}(1/2)$, so the numerator cannot be improved. Nevertheless, the denominator $H_{\infty}(Q)$ can be improved. Since $m$ copies of $Q$ can have approximately $mH(Q)$ amount of information that is ``nearly uniform'' by the asymptotic equipartition property, we expect the correct denominator to be $H(Q)$ instead of $H_{\infty}(Q)$. 

Indeed, the following theorem shows that $m^{*}(P\Vert Q)$ can be approximately upper-bounded by $H_{0}(P)/H(Q)$, at the expense of having a constant depending on $Q$ that cannot be easily tracked, making the bound only meaningful when $H_{0}(P)\gg H(Q)$.

\smallskip{}

\begin{thm}
\label{thm:H0_over_H}For every non-degenerate discrete distribution $Q$ with an information spectrum with finite third moment,\footnote{This means $\mathbb{E}[|\iota_{Y}(Y)|^{3}]<\infty$ when $Y\sim Q$, where $\iota_{Y}(y):=-\log Q(y)$ is the self-information.} there exists a constant $c_{Q}>0$ (which only depends on $Q$) such that
\[
m^{*}(P\Vert Q)\le\frac{H_{0}(P)}{H(Q)}+c_{Q}(H_{0}(P))^{2/3}
\]
for every discrete distribution $P$.
\end{thm}
\smallskip{}

As a result, we can characterize the first-order term of $m^{*}(P\Vert Q)$ when $P$ is uniform by comparing against the lower bound in Proposition \ref{prop:properties}. This shows that for a uniform source, the asymptotic number of source bits encoded per encoding symbol is $H(Q)$.

\smallskip{}

\begin{cor}
As $b\to\infty$, we have
\[
m^{*}(\mathrm{Unif}[b]\Vert Q)=\frac{\log b}{H(Q)}+O(\log^{2/3}b).
\]
Hence, $H(Q)=\lim_{b\to\infty}(\log b)/m^{*}(\mathrm{Unif}[b]\Vert Q)$.
\end{cor}
\smallskip{}

We now prove Theorem \ref{thm:H0_over_H}.
\begin{IEEEproof}
[Proof of Theorem \ref{thm:H0_over_H}]The case $H_{0}(P)=0$ is clear, so we assume $H_{0}(P)\ge1$ (it is impossible to have $0<H_{0}(P)<1$). Fix $k\in\mathbb{N}$, $k\ge7$. Write $H_{\infty}:=H_{\infty}(Q)$ and $H:=H(Q)=\mathbb{E}[\iota_{Y}(Y)]$ where $Y\sim Q$, and $\iota_{Y}(y):=-\log P_{Y}(y)$ is the self-information. Let $\sigma^{2}:=\mathrm{Var}(\iota_{Y}(Y))$. The result follows directly from Corollary \ref{thm:ratio} if $\sigma^{2}=0$, so we assume $\sigma^{2}>0$. Let $\mathbf{Y}=(Y_{1},\ldots,Y_{n})$, $Y_{1},\ldots,Y_{n}\stackrel{\mathrm{iid}}{\sim}Q$, $T:=(\iota_{\mathbf{Y}}(\mathbf{Y})-nH)/(\sigma\sqrt{n})$. Consider the inverse cumulative distribution function $F_{T}^{-1}$ of $T$, and $\Phi^{-1}$ of $\mathrm{N}(0,1)$.  Consider a quantile coupling between $T$ and $W\sim\mathrm{Unif}[0:k-1]$, i.e., we let $U\sim\mathrm{Unif}(0,1)$ be such that $T=F_{T}^{-1}(U)$, and let $W=\lfloor kU\rfloor$. We have $F_{T}(T)\ge W/k$. Since $\mathbb{E}[(\iota_{Y}(Y))^{3}]<\infty$, by the Berry-Esseen theorem \cite{berry1941accuracy,esseen1942liapunov}, $|F_{T}(t)-\Phi(t)|\le c/\sqrt{n}$, where $c>0$ is a constant that only depends on $Q$. Hence, $\Phi(T)\ge W/k-c/\sqrt{n}$, and 
\begin{align*}
T & \ge\Phi^{-1}(W/k-c/\sqrt{n})\\
 & \ge\Phi^{-1}(W/k)-\frac{c}{\sqrt{n}}\sup_{\tau\in[W/k-c/\sqrt{n},\,W/k]}\frac{\mathrm{d}\Phi^{-1}(\tau)}{\mathrm{d}\tau}\\
 & \ge\Phi^{-1}(W/k)-\frac{c/\sqrt{n}}{\inf_{t\in[\Phi^{-1}(W/k-c/\sqrt{n}),\,\Phi^{-1}(W/k)]}\phi(t)},
\end{align*}
where $\phi$ is the probability density function of $\mathrm{N}(0,1)$, and we take $\Phi^{-1}(\tau)=-\infty$ if $\tau\le0$ (we consider the above inequality to be vacuously true if $W/k-c/\sqrt{n}\le0$). Mills' inequality gives $1-\Phi(t)\le\phi(t)/t\le\phi(t)$ for $t\ge1$. Hence, $\phi(\Phi^{-1}(\tau))\ge1-\tau$ for $\tau\ge6/7$. Since $\phi(\Phi^{-1}(\tau))$ is decreasing in $|\tau-1/2|$, $\phi(\Phi^{-1}(\tau))\ge1/(2k)$ for $1/(2k)\le\tau\le1-1/(2k)$. Therefore, as long as $\sqrt{n}\ge2ck$, we have 
\begin{equation}
T\ge\Phi^{-1}(W/k)-\frac{2ck}{\sqrt{n}}.\label{eq:T_Phi_lower}
\end{equation}

 We generate $k$ copies of $\mathbf{Y}$, denoted as $\mathbf{Y}^{(0)},\ldots,\mathbf{Y}^{(k-1)}$, by generating $\mathbf{Y}^{(i)}$ conditional on $W$ following $P_{\mathbf{Y}|W}(\cdot|\,W+i\;\mathrm{mod}\;k)$. T Let $T^{(i)}:=(\iota_{\mathbf{Y}}(\mathbf{Y}^{(i)})-nH)/(\sigma\sqrt{n})$. By (\ref{eq:T_Phi_lower}), we have 
\[
T^{(i)}\ge\Phi^{-1}\Big(\frac{W+i\;\mathrm{mod}\;k}{k}\Big)-\frac{2ck}{\sqrt{n}}.
\]
We also have
\begin{align*}
 & P_{\mathbf{Y}^{(0)},\ldots,\mathbf{Y}^{(k-1)}}(\mathbf{y}^{(0)},\ldots,\mathbf{y}^{(k-1)})\\
 & =\frac{1}{k}\sum_{w=0}^{k-1}\prod_{i=0}^{k-1}P_{\mathbf{Y}|W}(\mathbf{y}^{(i)}|\,w+i\;\mathrm{mod}\;k)\\
 & \le\frac{1}{k}\sum_{w=0}^{k-1}\prod_{i=0}^{k-1}(kP_{\mathbf{Y}}(\mathbf{y}^{(i)}))\;=\;\prod_{i=0}^{k-1}(kP_{\mathbf{Y}}(\mathbf{y}^{(i)})).
\end{align*}
Hence, 
\begin{align*}
 & \iota_{\mathbf{Y}^{(0)},\ldots,\mathbf{Y}^{(k-1)}}(\mathbf{Y}^{(0)},\ldots,\mathbf{Y}^{(k-1)})+k\log k\\
 & \ge\sum_{i=0}^{k-1}\iota_{\mathbf{Y}}(\mathbf{Y}^{(i)})\\
 & =\sum_{i=0}^{k-1}\max\left\{ \sigma\sqrt{n}T^{(i)}+nH,\,0\right\} \\
 & \ge\sum_{i=0}^{k-1}\max\left\{ \sigma\sqrt{n}\Big(\Phi^{-1}\Big(\frac{W+i\;\mathrm{mod}\;k}{k}\Big)-\frac{2ck}{\sqrt{n}}\Big)+nH,\,0\right\} \\
 & \ge\sum_{i=1}^{k-1}\Big(\sigma\sqrt{n}\Big(\Phi^{-1}(i/k)-\frac{2ck}{\sqrt{n}}\Big)+nH\Big)\\
 & \ge n(k-1)H+\sigma\sqrt{n}\sum_{i=1}^{k-1}\Phi^{-1}(i/k)-2ck^{2}\sigma\\
 & =n(k-1)H-2ck^{2}\sigma.
\end{align*}
Hence, 
\begin{align}
H_{\infty}(\mathbf{Y}^{(0)},\ldots,\mathbf{Y}^{(k-1)}) &\ge\zeta:=n(k-1)H-(2c\sigma+1)k^{2}. \label{eq:pf_H_infty}
\end{align}
Taking $n=\lceil(2ck)^{2}\rceil$, we have $\zeta \ge 4c^2 H k^2(k - c')$, where $c' := 1+(2c \sigma + 1)/(4c^2H)$.
Note that $\mathbf{Y}^{(0)},\ldots,\mathbf{Y}^{(k-1)}$ is a coupling of $nk$ copies of $Q$,
and their joint distribution is majorized by $P$ if $\zeta \ge \alpha := H_0(P)$.
Taking $k=\lceil (4c^2H/\alpha)^{-1/3} + c' \rceil$, 
we have $\zeta \ge 4c^2 H k^2(k - c') \ge \alpha$.
Applying Lemma \ref{lem:k_plus_2}, for large $\alpha$,
\begin{align*}
m^{*}(P\Vert Q) & \le nk + \max\Big\{\Big\lceil\frac{1/2}{1-2^{-H_{\infty}}}\Big\rceil, 2\Big\} \\
 & \le (2ck)^{2}k + k +\frac{1}{H_{\infty}} + 2\\
 & = 4c^2 ((4c^2H/\alpha)^{-1/3} + O(1))^3 + O(\alpha^{1/3}) \\
 & = \alpha/H + O(\alpha^{2/3}),
\end{align*}
in the sense that there exists $\alpha_{0},c''$ (which depend on $Q$) such that $m^{*}(P\Vert Q)\le\alpha/H+c''\alpha^{2/3}$ for $\alpha\ge\alpha_{0}$. For $1\le\alpha<\alpha_{0}$, Theorem \ref{thm:ratio} gives $m^{*}(P\Vert Q)<(\alpha_{0}+1)/H_{\infty}+3$. The result follows from taking $c_{Q}:=\max\{c'',(\alpha_{0}+1)/H_{\infty}+3\}$.


\end{IEEEproof}
\smallskip{}

\begin{remark}
The inequality \eqref{eq:pf_H_infty} implies the following result:
\[
\lim_{m\to \infty}\frac{1}{m} \sup_{\bar{Q}\in \Gamma_m(Q)}H_\infty(\bar Q) = H(Q).
\]
Comparing this to $(1/m) H_{\infty}(Y_1,\ldots,Y_m) = H_{\infty}(Q)$ when $Y_1,\ldots,Y_m \stackrel{\mathrm{iid}}{\sim} Q$, we can see that relaxing the i.i.d. condition to the requirement that the variables be identically distributed can increase the min-entropy of the coupling.
\end{remark}

\section{Asymptotic SC-MDC}

We now consider the asymptotic SC-MDC setting where the source $X^{n}\sim P^{\otimes n}$ is i.i.d. and $n\to\infty$. Define the \emph{asymptotic coupled aggregation ratio} as
\begin{equation}
\gamma(P\Vert Q):=\lim_{n\to\infty}\frac{1}{n}m^{*}(P^{\otimes n}\Vert Q).\label{eq:asymp_ratio}
\end{equation}
Note that the limit exists by Fekete's lemma since $m^{*}(P^{\otimes n}\Vert Q)$ is subadditive in $n$. We can exactly characterize $\gamma(P\Vert Q)$ when $P$ is finite, showing that the lower bound (\ref{eq:prop_lower_bound}) is asymptotically tight.

\smallskip{}

\begin{thm}
\label{thm:gamma}For discrete distributions $P,Q$ with $H_{0}(P)<\infty$,  we have
\begin{equation}
\gamma(P\Vert Q)=\max\Big\{\frac{H(P)}{H(Q)},\frac{H_{0}(P)}{H_{0}(Q)}\Big\}.\label{eq:thm_gamma}
\end{equation}
We regard $0/0=0$ and $a/0=\infty$ (where $a>0$) here.
\end{thm}
\smallskip{}

An interesting special case is $H_{0}(Q)=\infty$, where $\gamma(P\Vert Q)=H(P)/H(Q)$. For example, we have 
\begin{equation}
\gamma(P\Vert\mathrm{Geom}(a))=aH(P)/H_{\mathrm{b}}(a)\label{eq:geom}
\end{equation}
for $0<a<1$, where $H_{\mathrm{b}}$ is the binary entropy function. This has the following operational interpretation. Suppose we want to encode $X^{n}\sim P^{\otimes n}$ into $Y_{1},\ldots,Y_{m}\in\mathbb{Z}_{\ge0}$, where we have a cost constraint $\mathbb{E}[Y_{i}]\le1/a-1$ for $i\in[m]$. In other words, we find a joint distribution of $(X^{n},Y^{m})$ with $X^{n}\sim P^{\otimes n}$, $\mathbb{E}[Y_{i}]\le1/a-1$ and $H(X^{n}|Y^{m})=0$. We say that a rate $R>0$ is achievable if the encoding is possible with $m=\lfloor nR\rfloor$ for all large enough $n$. We can show that the infimum of achievable $R$ is $aH(P)/H_{\mathrm{b}}(a)$. The achievability follows from (\ref{eq:geom}) by taking $Y_{i}\sim\mathrm{Geom}(a)$,\footnote{We consider $\mathrm{Geom}(a)$ to have a support $\mathbb{Z}_{\ge0}$ instead of $\mathbb{Z}_{>0}$.} whereas the converse follows from $H(X^{n})\le H(Y^{m})\le\sum_{i}H(Y_{i})\le mH_{\mathrm{b}}(a)/a$ since the geometric distribution maximizes entropy for a given expectation. 

In particular, the optimal $R$ for $a=1/2$ is $H(P)/2$, giving a new operational meaning of entropy. Comparing this to the standard operational meaning of entropy via source coding, which is the rate needed to encode $X^{n}$ to a fixed number of bits with vanishing error probability, or to a variable number of bits (via a prefix code) with zero error \cite{shannon1948mathematical}, our new operational interpretation is fixed-length and zero error, but requires a cost constraint. This is perhaps surprising since it is usually expected  that to achieve a rate $H(P)$ (instead of $H_{0}(P)$) in source coding, we either require a non-zero vanishing error probability, or a variable-length code.

\smallskip{}

\begin{IEEEproof}
[Proof of Theorem \ref{thm:gamma}]If $H(Q)=0$, we clearly have $\gamma(P\Vert Q)=0$ if $H(P)=0$, or $\gamma(P\Vert Q)=\infty$ otherwise, so we assume $H(Q)>0$. The ``$\ge$'' direction follows from the fact that if $X^{n}\sim P^{\otimes n}$, $Y_{1},\ldots,Y_{m}\sim Q$ with $H(X^{n}|Y^{m})=0$, then $nH(X)=H(X^{n})\le H(Y^{m})\le mH(Y)$, and the same holds with $H_{0}$ in place of $H$. We now prove the ``$\le$'' direction. We first consider the case $H_{0}(Q)<\infty$.  Consider any integer $k\le|\mathrm{supp}(Q)|$, and let $Q_{1}\ll Q$ be a distribution that is uniform over a support of size $k$. For every $0<\epsilon<k\min_{y\in\mathrm{supp}(Q_{1})}Q(y)$, we can decompose $Q$ into a mixture $Q=(1-\epsilon)Q_{0}+\epsilon Q_{1}$, with $H(Q_{0})\to H(Q)$ as $\epsilon\to0$. Assume $X^{n}\sim P^{\otimes n}$. By the law of large numbers, for every large enough $n$,
\begin{align}
& \mathbb{P}(\iota_{X^{n}}(X^{n})>(1+\epsilon)nH(P))\le\epsilon. \label{eq:pf_info_stable}
\end{align}
Hence, we can define an event $E$ with $\mathbb{P}(E)=1-\epsilon$ such that if $E$ occurs, then $\iota_{X^{n}}(X^{n})\le(1+\epsilon)nH(P)$ (note that $E$ might require additional randomness outside of $X^{n}$).  By almost quasiconvexity (Proposition \ref{prop:properties}), 
\begin{align*}
 & \frac{1}{n}m^{*}(P^{\otimes n}\Vert Q)\\
 & \le\frac{1}{n}\max\left\{ m^{*}(P_{X^{n}|E}\big\Vert Q_{0}),m^{*}(P_{X^{n}|E^{\mathrm{c}}}\big\Vert Q_{1})\right\} +\frac{1}{n}m^{*}(\mathrm{Bern}(\epsilon)\Vert Q)\\
 & \stackrel{(a)}{\le}\frac{1}{n}\max\bigg\{\frac{H_{0}(P_{X^{n}|E})}{H(Q_{0})}+c_{Q_{0}}\big((H_{0}(P_{X^{n}|E}))^{2/3}+1\big),\\
 & \;\quad\frac{H_{0}(P_{X^{n}|E^{\mathrm{c}}})+1}{H_{\infty}(Q_{1})}\bigg\}+\frac{2}{nH_{\infty}(Q)}+\frac{6}{n}\\
 & \stackrel{(b)}{\le}\max\bigg\{\frac{(1+\epsilon)H(P)}{H(Q_{0})}+c_{Q_{0}}\big(((1+\epsilon)H(P))^{2/3}n^{-1/3}+1/n\big),\\
 & \;\quad\frac{H_{0}(P)+1/n}{\log k}\bigg\}+\frac{2}{nH_{\infty}(Q)}+\frac{6}{n}\\
 & \to\max\left\{ \frac{H(P)}{H(Q)},\,\frac{H_{0}(P)}{\log k}\right\},
\end{align*}
as $n\to\infty$ and then $\epsilon\to0$, where (a) is by Theorem \ref{thm:H0_over_H} (the finite third moment assumption in Theorem \ref{thm:H0_over_H} is satisfied since $Q_0$ is finite) and Corollary \ref{thm:ratio}, and (b) is because there are at most $2^{(1+\epsilon)nH(X)}$ choices of $X^{n}$ where the event $E$ holds. The result follows from taking $k=|\mathrm{supp}(Q)|$ if $|\mathrm{supp}(Q)|<\infty$, or $k\to\infty$ otherwise.

If $H_{0}(Q)=\infty$ and $H(Q)<\infty$, we can truncate $Q$ to a finite $\hat{Q}$ with $H_{0}(\hat{Q})>1/\epsilon$ and $H(\hat{Q})>H(Q)-\epsilon$, and apply the finite case on $\hat{Q}$. The result follows from taking $\epsilon\to0$. If $H_{0}(Q)=H(Q)=\infty$, we can truncate $Q$ to a finite $\hat{Q}$ with $H_{0}(\hat{Q})\ge H(\hat{Q})>1/\epsilon$. The result follows from taking $\epsilon\to0$. 
\end{IEEEproof}
\smallskip{}

Note that the proof of Theorem \ref{thm:gamma} also applies to non-i.i.d. $X^n$, as long as \eqref{eq:pf_info_stable} is satisfied. A similar bound on $\iota_{X^n}(X^n)$ also holds if $\mathbf{X}=(X_1,X_2,\ldots)$ is information stable \cite{pinsker1964information} (which holds, for example, when $\mathbf{X}$ is stationary and ergodic with finite entropy rate \cite{pinsker1964information}).
\begin{thm}
\label{thm:process}For a discrete distribution $Q$ with $H(Q)>0$, and a source $\mathbf{X}=(X_1,X_2,\ldots)$ where $\bar{H}(\mathbf{X}) = \lim_{n \to \infty}n^{-1} H(X^n)$ and $\bar{H}_0(\mathbf{X}) = \lim_{n \to \infty}n^{-1} H_0(X^n)$ exist and are finite, and $\iota_{X^n}(X^n)/n$ converges to $\bar{H}(\mathbf{X})$ in probability, we have
\begin{equation}
\lim_{n \to \infty}\frac{1}{n} m^*(P_{X^n} \Vert Q)=\max\Big\{\frac{\bar{H}(\mathbf{X})}{H(Q)},\frac{\bar{H}_0(\mathbf{X})}{H_{0}(Q)}\Big\}.\label{eq:thm_gamma}
\end{equation}
\end{thm}

\begin{IEEEproof}
Since $\iota_{X^n}(X^n)/n$ converges to $\bar{H}(\mathbf{X})$ in probability, we have $\mathbb{P}(\iota_{X^{n}}(X^{n})>n(\bar{H}(\mathbf{X})+\epsilon))\le\epsilon$ for $n$ large enough. The rest of the proof follows from the same arguments as in the proof of Theorem \ref{thm:gamma}, with $\bar{H}(\mathbf{X})+\epsilon$ in place of $(1+\epsilon)H(P)$.
\end{IEEEproof}

The characterization of $\gamma(P\Vert Q)$ for $P,Q$ both having infinite support is left for future studies. This will likely depend on the tail behaviors of $P$ and $Q$.

\smallskip{}

\section{Approximate Marginal Constraint and the Rate-Perception Function}\label{sec:perception}

In this section, we generalize the SC-MDC setting by relaxing the marginal constraint to only require $P_{Y_{i}}\approx Q$. More precisely, we require $\delta(Q,P_{Y_{i}})\le\Delta$ for all $i\in[m]$, where $\delta:\mathcal{P}(\mathcal{Y})\times\mathcal{P}(\mathcal{Y})\to[0,\infty]$ is a certain function (where $\mathcal{P}(\mathcal{Y})$ is the set of discrete distributions over $\mathcal{Y}$), and $\Delta\ge0$ is the allowed divergence level. If we regard $P_{Y_{i}}=Q$ as $Y_{i}$ having perfect perceptual quality as discussed in the introduction, and if $\delta$ is a divergence, then requiring $\delta(Q,P_{Y_{i}})\le\Delta$ (or equivalently, requiring $P_{Y_i} \in \mathcal{A} := \{Q': \delta(Q,Q')\le\Delta\}$) is a constraint that guarantees high perceptual quality, in line with the rate-distortion-perception function \cite{blau2018perception,blau2019rethinking,theis2021coding,chen2022rate}. Similar symbol-wise perception constraints are imposed in the rate-distortion-perception coding theorems in \cite{chen2022rate,theis2021coding}.

Another setting where we do not require an exact marginal distribution is source coding with symbol-wise expected cost constraints. Suppose we require that $\mathbb{E}[c_j(Y_i)] \le w_j$, $i\in[m]$, $j\in[\ell]$, for cost functions $c_1,\ldots,c_\ell : \mathcal{Y} \to \mathbb{R}$ and $w_1,\ldots,w_\ell \in \mathbb{R}$. A motivation is wear leveling as discussed in the introduction, where each $c_j$ corresponds to a degradation mechanism (a way in which the memory cell may degrade). We impose the constraint $P_{Y_i} \in \mathcal{A} := \{Q': \forall j.\, \mathbb{E}_{Y \sim Q'}[c_j(Y)] \le w_j\}$.

Generally, we can impose the constraint $P_{Y_i} \in \mathcal{A}$ for any set $\mathcal{A} \subseteq \mathcal{P}(\mathcal{Y})$.
The optimal rate is defined as follows.

\smallskip{}

\begin{defn}
Given a distribution $P$ and set $\mathcal{A} \subseteq \mathcal{P}(\mathcal{Y})$, the \emph{asymptotic coupled aggregation ratio between $P$ and $\mathcal{A}$} is defined as\footnote{The limit exists by Fekete's lemma since $m^{*}(P^{\otimes n}\Vert \mathcal{A})$ is subadditive in $n$.}
\[
\gamma(P\Vert \mathcal{A}):=\lim_{n\to\infty}\frac{1}{n}m^{*}(P^{\otimes n}\Vert \mathcal{A}),
\]
where
\begin{align*}
m^{*}(R\Vert \mathcal{A}) & :=\min\big\{ m\in\mathbb{Z}_{\ge0}:\,\exists Q_{1},\ldots,Q_{m}\in\mathcal{A}.\\
 & \quad\quad\quad\quad\; (Q_{1},\ldots,Q_{m})\sqsubseteq R \big\}.
\end{align*}
In particular, given distributions $P,Q$, function $\delta:\mathcal{P}(\mathcal{Y})\times\mathcal{P}(\mathcal{Y})\to[0,\infty]$ and $\Delta\ge0$, the \emph{SC-MDC rate-perception function} is 
\[
\gamma_{\Delta}(P\Vert Q):=\gamma(P \,\Vert\, \{Q': \delta(Q,Q')\le\Delta\}).
\]
\end{defn}
\smallskip{}

We now characterize the asymptotic coupled aggregation ratio between $P$ and $\mathcal{A}$.

\smallskip{}

\begin{thm}
\label{thm:perception}For a discrete distribution $P$ with $H_{0}(P)<\infty$ and nonempty $\mathcal{A} \subseteq \mathcal{P}(\mathcal{Y})$, we have
\begin{align}
\gamma(P\Vert \mathcal{A}) & =\inf_{\hat{Q}_{0},\hat{Q}_{1} \in \mathcal{A},\,\lambda\in[0,1]}\nonumber \\
 & \quad\quad\max_{\alpha\in\{0,1\}}\frac{H_{\alpha}(P)}{(1-\lambda)H_{\alpha}(\hat{Q}_{0})+\lambda H_{\alpha}(\hat{Q}_{1})},\label{eq:thm_perception}
\end{align}
where $H_{1}(P)=H(P)$. Moreover, if $\mathcal{A}$ is convex, this can be simplified to 
\begin{equation}
\gamma(P\Vert \mathcal{A}) =\max_{\alpha\in\{0,1\}}\frac{H_{\alpha}(P)}{\sup_{\hat{Q} \in \mathcal{A}}H_{\alpha}(\hat{Q})}.\label{eq:thm_perception_conv}
\end{equation}
\end{thm}

\smallskip{}

For a divergence $\delta$ that is convex in its second argument, $\mathcal{A} = \{Q': \delta(Q,Q')\le\Delta\}$ is convex, and hence we can find the rate-perception function $\gamma_{\Delta}(P\Vert Q)$ via \eqref{eq:thm_perception_conv} by solving $\sup_{\hat{Q}:\delta(Q,\hat{Q})\le\Delta}H_{\alpha}(\hat{Q})$ for $\alpha \in \{0,1\}$.
The optimization problem $\sup_{\hat{Q}:\delta(Q,\hat{Q})\le\Delta}H_{0}(\hat{Q})$ is usually simple. For example, if $\delta(Q,Q')=D_{\mathrm{KL}}(Q'\Vert Q)$, then the optimal value of the optimization problem is $H_{0}(Q)$. If $\delta(Q,Q')=d_{\mathrm{TV}}(Q,Q')$ or $D_{\mathrm{KL}}(Q\Vert Q')$, then the optimal value is $\log|\mathcal{Y}|$ as long as $\Delta>0$. Hence, when $\delta$ is a divergence that is convex in its second argument, to find $\gamma_{\Delta}(P\Vert Q)$, we usually only need to solve $\sup_{\hat{Q}:\delta(Q,\hat{Q})\le\Delta}H(\hat{Q})$ which is a convex optimization problem. The following results follow from straightforward applications of the Karush-Kuhn-Tucker conditions.

\smallskip{}

\begin{cor}
Under the same assumptions as Theorem \ref{thm:perception}, we have:
\begin{itemize}
\item If $\delta(Q,Q')=D_{\mathrm{KL}}(Q'\Vert Q)$, then
\[
\gamma_{\Delta}(P\Vert Q)=\max\Big\{\frac{H(P)}{H(Q^{\lambda})},\frac{H_{0}(P)}{H_{0}(Q)}\Big\},
\]
where $Q^{\lambda}$ is the tilted distribution \cite{polyanskiy2024information} with $Q^{\lambda}(y)\propto(Q(y))^{\lambda}$ (we take $0^{0}=0$ here), for $\lambda=\min\{\lambda\in[0,1]:\,D_{\mathrm{KL}}(Q^{\lambda}\Vert Q)\le\Delta\}$.
\item If $\delta(Q,Q')=d_{\mathrm{TV}}(Q,Q')$ and $\Delta>0$ and $\mathcal{Y}$ is finite, then
\[
\gamma_{\Delta}(P\Vert Q)=\max\Big\{\frac{H(P)}{H(Q_{[a,b]})},\frac{H_{0}(P)}{\log|\mathcal{Y}|}\Big\},
\]
where $Q_{[a,b]}(y):=\min\{\max\{Q(y),a\},b\}$ for $0\le a\le b\le1$ chosen such that $\sum_{y}Q_{[a,b]}(y)=1$, $d_{\mathrm{TV}}(Q,Q_{[a,b]})\le\Delta$, and ($a=b$ or $d_{\mathrm{TV}}(Q,Q_{[a,b]})=\Delta$). If $\mathcal{Y}$ is infinite and $\Delta>0$, $\gamma_{\Delta}(P\Vert Q) = 0$.\footnote{If $\mathcal{Y}$ is infinite, we can find $Q'$ with $d_{\mathrm{TV}}(Q,Q') \le \Delta$, $H(Q')=\infty$ and $H_0(Q')=\infty$.}
\end{itemize}
\end{cor}
\smallskip{}

We now prove Theorem \ref{thm:perception}.
\begin{IEEEproof}
[Proof of Theorem \ref{thm:perception}]We first prove the achievability. Fix $\hat{Q}_{0},\hat{Q}_{1} \in \mathcal{A}$ and $\lambda\in[0,1]$. Fix $\epsilon>0$, $k\ge1$, and let $Q=\hat{Q}_{0}^{\otimes(k-\lfloor\lambda k\rfloor)}\otimes\hat{Q}_{1}^{\otimes\lfloor\lambda k\rfloor}$. Applying Theorem \ref{thm:gamma} on $Q$, we have 
\begin{align*}
\gamma(P\Vert Q) & =\max_{\alpha\in\{0,1\}}\frac{H_{\alpha}(P)}{(k-\lfloor\lambda k\rfloor)H_{\alpha}(\hat{Q}_{0})+\lfloor\lambda k\rfloor H_{\alpha}(\hat{Q}_{1})}\\
 & \le(1+\epsilon)\zeta/k,
\end{align*}
for $k$ large enough, where
\[
\zeta:=\max_{\alpha\in\{0,1\}}\frac{H_{\alpha}(P)}{(1-\lambda)H_{\alpha}(\hat{Q}_{0})+\lambda H_{\alpha}(\hat{Q}_{1})}.
\]
Therefore, $P^{\otimes n}$ is a coupled aggregation of $(k-\lfloor\lambda k\rfloor)n(1+\epsilon)\zeta/k+o(n)$ copies of $\hat{Q}_{0}$ and $\lfloor\lambda k\rfloor n(1+\epsilon)\zeta/k+o(n)$ copies of $\hat{Q}_{1}$. We have $m^{*}(P^{\otimes n}\Vert \mathcal{A})\le n(1+\epsilon)\zeta+o(n)$, and hence $\gamma(P\Vert \mathcal{A})\le(1+\epsilon)\zeta$. Taking $\epsilon\to0$ shows that $\gamma(P\Vert \mathcal{A})$ is upper-bounded by the right-hand side of (\ref{eq:thm_perception}).

We then prove the converse. Assume there exists $Q_{1},\ldots,Q_{m} \in \mathcal{A}$ with $(Q_{1},\ldots,Q_{m})\sqsubseteq P^{\otimes n}$. We have $nH_{\alpha}(P)\le\sum_{i=1}^{m}H_{\alpha}(Q_{i})$ for $\alpha\in\{0,1\}$. The point $(h_{0},h_{1}):=(m^{-1}\sum_{i=1}^{m}H_{0}(Q_{i}),m^{-1}\sum_{i=1}^{m}H(Q_{i}))\in\mathbb{R}^{2}$ is in the convex hull of $\bigcup_{i=1}^{m}\{(H_{0}(Q_i),H(Q_i))\}$. By Carath\'{e}odory's theorem, there exist $\hat{Q}_{0},\hat{Q}_{1} \in \mathcal{A}$ and $\lambda\in[0,1]$ such that $h_{\alpha}\le(1-\lambda)H_{\alpha}(\hat{Q}_{0})+\lambda H_{\alpha}(\hat{Q}_{1})$ for $\alpha\in\{0,1\}$.\footnote{Carath\'{e}odory's theorem shows that $(h_{0},h_{1})$ is the convex combination of $3$ points. We need only $2$ points since we only require upper bounds on $h_{0},h_{1}$. This can be shown by noting that any triangle is a subset of the downward closure of its three edges.} Therefore, $H_{\alpha}(P)\le(m/n)((1-\lambda)H_{\alpha}(\hat{Q}_{0})+\lambda H_{\alpha}(\hat{Q}_{1}))$, and $m/n$ is lower-bounded by the right-hand side of (\ref{eq:thm_perception}). Hence, $\gamma(P\Vert \mathcal{A})$ is also lower-bounded by the same quantity.

We finally consider the case where $\mathcal{A}$ is convex. Since $\min\max\ge\max\min$, the right-hand side of (\ref{eq:thm_perception}) is lower-bounded by the right-hand side of (\ref{eq:thm_perception_conv}). For the other direction, let $0<\epsilon<1$, and consider any $\hat{Q}_{0},\hat{Q}_{1} \in \mathcal{A}$ with $H_{\alpha}(\hat{Q}_{\alpha})\ge(1-\epsilon)h_{\alpha}^{*}$ where $h_{\alpha}^{*}:=\sup_{\hat{Q} \in \mathcal{A}}H_{\alpha}(\hat{Q})$. Let $\hat{Q}=\epsilon\hat{Q}_{0}+(1-\epsilon)\hat{Q}_{1} \in \mathcal{A}$ since $\mathcal{A}$ is a convex set. We have
\begin{align*}
 & \max\Big\{\frac{H(P)}{H(\hat{Q})},\frac{H_{0}(P)}{H_{0}(\hat{Q})}\Big\}\\
 & \le\max\Big\{\frac{H(P)}{\epsilon H(\hat{Q}_{0})+(1-\epsilon)H(\hat{Q}_{1})},\frac{H_{0}(P)}{\max\{H_{0}(\hat{Q}_{0}),H_{0}(\hat{Q}_{1})\}}\Big\}\\
 & \le\max\Big\{\frac{H(P)}{(1-\epsilon)H(\hat{Q}_{1})},\frac{H_{0}(P)}{H_{0}(\hat{Q}_{0})}\Big\}\\
 & \le\frac{1}{(1-\epsilon)^{2}}\max\Big\{\frac{H(P)}{h_{1}^{*}},\frac{H_{0}(P)}{h_{0}^{*}}\Big\}.
\end{align*}
Taking $\epsilon\to0$, the right-hand side of (\ref{eq:thm_perception}) is upper-bounded by the right-hand side of (\ref{eq:thm_perception_conv}).
\end{IEEEproof}

\smallskip{}

\section{SC-MDC with Vanishing Error}

We can also consider an asymptotic SC-MDC setting with i.i.d. source $X^{n}\sim P^{\otimes n}$, where we only require the decoding error probability to be small instead of zero. More precisely, we design $P_{Y^{m}|X^{n}}$ and a decoding function $g:\mathcal{Y}^{m}\to\mathcal{X}^{n}$ such that $Y_{i}\sim Q$ and $\mathbb{P}(g(Y^{m})\neq X^{n})\le\epsilon$. This is equivalent to finding a distribution $\hat{P}$ over $\mathcal{X}^{n}$ with total variation distance $d_{\mathrm{TV}}(P^{\otimes n},\hat{P})\le\epsilon$ and $m^{*}(\hat{P}\Vert Q)\le m$. To prove the forward direction, we can take $\hat{P}$ to be the distribution of $g(Y^{m})$, so $d_{\mathrm{TV}}(P^{\otimes n},\hat{P})\le\mathbb{P}(g(Y^{m})\neq X^{n})\le\epsilon$. For the reverse direction, if we have such a $\hat{P}$, we can construct a scheme for vanishing-error SC-MDC by letting $\hat{X}^{n}\sim\hat{P}$ and $Y_{i}\sim Q$ be coupled so that $H(\hat{X}^{n}|Y^{m})=0$, construct $X^{n}\sim P^{\otimes n}$ which is maximally coupled with $\hat{X}^{n}$, i.e., $\mathbb{P}(X^{n}\neq\hat{X}^{n})=d_{\mathrm{TV}}(P^{\otimes n},\hat{P})\le\epsilon$, and take the stochastic encoder $P_{Y^{m}|X^{n}}$ to be the conditional distribution of $Y^{m}$ given $X^{n}$. 

We can show that the optimal rate for vanishing-error SC-MDC is $H(P)/H(Q)$. Comparing to Theorem \ref{thm:gamma}, we can see that allowing a vanishing error removes the dependency on the Hartley entropies $H_{0}(P),H_{0}(Q)$.

\smallskip{}

\begin{thm}
\label{thm:vanishing_error}For discrete distributions $P,Q$ with $H_{0}(P)<\infty$, 
\begin{equation}
\lim_{\epsilon\to0}\underset{n\to\infty}{\mathrm{limsup}}\frac{1}{n}\inf_{\hat{P}:d_{\mathrm{TV}}(P^{\otimes n},\hat{P})\le\epsilon}m^{*}(\hat{P}\Vert Q)=\frac{H(P)}{H(Q)},\label{eq:thm_approx}
\end{equation}
where the infimum is over discrete distributions $\hat{P}$ over $\mathcal{X}^{n}$.
\end{thm}
\smallskip{}
\begin{IEEEproof}
Compared to Theorem \ref{thm:gamma}, Theorem \ref{thm:vanishing_error} does not depend on $H_0(P)$. Intuitively, allowing a vanishing error lets us focus only on typical sequences of $X^n$, and remove the dependency on the cardinality of $P$. We now prove the achievability.
We use the same proof as Theorem \ref{thm:gamma} for the case $H_{0}(Q)<\infty$, except that we define $\hat{P}:=P_{X^{n}|E}$, $\hat{X}^{n}:=X^{n}$ if the event $E$ occurs, and $\hat{X}^{n}$ is generated from $\hat{P}$ (without depending on other random variables) if $E$ does not occur. Note that $\hat{X}^{n}\sim\hat{P}$.  We have $\mathbb{P}(\hat{X}^{n}\neq X^{n})\le1-\mathbb{P}(E)\le\epsilon$, so $d_{\mathrm{TV}}(P^{\otimes n},\hat{P})\le\epsilon$.  By the same arguments as in the proof of Theorem \ref{thm:gamma}, 
\begin{align*}
 & n^{-1} m^{*}(\hat{P}\big\Vert Q)\\
 & \le\frac{(1+\epsilon)H(P)}{H(Q)}+c_{Q}\big(((1+\epsilon)H(P))^{2/3}n^{-1/3}+1/n\big).
\end{align*}
Hence, $\mathrm{limsup}_{n}n^{-1}\inf_{\hat{P}:d_{\mathrm{TV}}(P^{\otimes n},\hat{P})\le\epsilon}m^{*}(\hat{P}\Vert Q)\le(1+\epsilon)H(P)/H(Q)$, and the left-hand side of (\ref{eq:thm_approx}) is at most $H(P)/H(Q)$.
For the case $H_{0}(Q)=\infty$, we use the same truncation argument as in the proof of Theorem \ref{thm:gamma}.

For the converse, letting $\hat{X}^{n}\sim\hat{P}$ and $Y_{i}\sim Q$ be coupled so that $H(\hat{X}^{n}|Y^{m})=0$, construct $X^{n}\sim P^{\otimes n}$ which is maximally coupled with $\hat{X}^{n}$, i.e., $\mathbb{P}(X^{n}\neq\hat{X}^{n})=d_{\mathrm{TV}}(P^{\otimes n},\hat{P})\le\epsilon$. By Fano's inequality \cite{fano1961transmission}, $H(X^{n})\le H(Y^{m})+\epsilon H_{0}(X^{n})+1$, so $m\ge(nH(P)-n\epsilon H_{0}(P)-1)/H(Q)$. Hence, $\mathrm{limsup}_{n}n^{-1}\inf_{\hat{P}:d_{\mathrm{TV}}(P^{\otimes n},\hat{P})\le\epsilon}m^{*}(\hat{P}\Vert Q)\ge(H(P)-\epsilon H_{0}(P))/H(Q)$, and the left-hand side of (\ref{eq:thm_approx}) is at least $H(P)/H(Q)$.
\end{IEEEproof}
\smallskip{}

We can combine Theorems \ref{thm:perception} and \ref{thm:vanishing_error} to obtain the following result on approximate SC-MDC with vanishing error. 
\begin{thm}
\label{thm:approx_vanishing_error}For a discrete distribution $P$ with $H_{0}(P)<\infty$ and nonempty $\mathcal{A} \subseteq \mathcal{P}(\mathcal{Y})$, we have
\begin{align}
&\lim_{\epsilon\to0}\underset{n\to\infty}{\mathrm{limsup}}\frac{1}{n}\inf_{\hat{P}:d_{\mathrm{TV}}(P^{\otimes n},\hat{P})\le\epsilon}m^{*}(\hat{P}\Vert \mathcal{A})  = \frac{H(P)}{\sup_{\hat{Q} \in \mathcal{A}}H(\hat{Q})}. \label{eq:approx_vanishing_error}
\end{align}
\end{thm}
\begin{IEEEproof}
For an upper bound, since $m^{*}(\hat{P}\Vert \mathcal{A}) \le m^{*}(\hat{P}\Vert \hat{Q})$ for every $\hat{Q} \in \mathcal{A}$, the left-hand side of \eqref{eq:approx_vanishing_error} is upper-bounded by $H(P)/H(\hat{Q})$ for every $\hat{Q} \in \mathcal{A}$ by Theorem \ref{thm:vanishing_error}. For a lower bound, we use the same converse proof as Theorem \ref{thm:vanishing_error}, except that $Y_i \sim Q_i$ where $Q_i \in \mathcal{A}$ for $i \in [m]$. We have $H(Y^m) \le m \sup_{\hat{Q} \in \mathcal{A}}H(\hat{Q})$.
\end{IEEEproof}

\section{Axiomatic Characterization of Shannon and Hartley Entropy}

In this section, we investigate a slightly unexpected consequence of Theorem \ref{thm:gamma} outside of the SC-MDC setting. We will show that Theorem \ref{thm:gamma} gives a new axiomatic characterization of Shannon and Hartley entropy that is similar to \cite{aczel1974shannon}. Consider a function $\tilde{H}$ from the space of finite discrete probability distributions to real numbers. For $\tilde{H}$ to be a reasonable entropy measure, we may expect $\tilde{H}$ to satisfy the following properties for finite discrete random variables (write $\tilde{H}(X)=\tilde{H}(P_{X})$):
\begin{itemize}
\item (Monotonicity) $\tilde{H}(X)\ge\tilde{H}(f(X))$ for every function $f$.
\item (Additivity) If $X,Y$ are independent, then $\tilde{H}(X,Y)=\tilde{H}(X)+\tilde{H}(Y)$. 
\item (Quasi-subadditivity) $\tilde{H}(X,Y)\le2\max\{\tilde{H}(X),\tilde{H}(Y)\}$.
\item (ID-subadditivity) If $X_{1},\ldots,X_{n}$ are identically distributed, $n\ge2$, then $\tilde{H}(X_{1},\ldots,X_{n})\le n\tilde{H}(X_{1})$.     
\end{itemize}
Note that quasi-subadditivity and ID-subadditivity are weak versions of the subadditivity property $\tilde{H}(X,Y)\le\tilde{H}(X)+\tilde{H}(Y)$. We can show that, as long as $\tilde{H}$ satisfies monotonicity, additivity, and either one of quasi-subadditivity or ID-subadditivity, then $\tilde{H}$ must be a non-negative linear combination of Shannon and Hartley entropy. Compare this to the celebrated result in \cite{aczel1974shannon}, which shows that as long as $\tilde{H}$ satisfies monotonicity,\footnote{Actually, \cite{aczel1974shannon} only requires $\tilde{H}(X)\ge\tilde{H}(f(X))$ for $f$ that is injective over the support of $X$.} additivity and weak subadditivity (i.e., $\tilde{H}(X,Y)\le \tilde{H}(X)+ \tilde{H}(Y)$ if $Y\in\{0,1\}$), then $\tilde{H}$ must be a non-negative linear combination of Shannon and Hartley entropy. Our result shows that we can replace weak subaddivity in \cite{aczel1974shannon} by either quasi-subadditivity or ID-subadditivity, and the result still holds. Also refer to \cite{csiszar2008axiomatic,muller2016generalization,baez2011characterization,li2023characterization} for related axiomatic characterizations.

\smallskip{}

\begin{thm}
If $\tilde{H}$ satisfies monotonicity+additivity+quasi-subadditivity, or satisfies monotonicity+additivity+ID-subadditivity, then $\tilde{H}$ must be of the form
\[
\tilde{H}(P)=\alpha H(P)+\beta H_{0}(P)
\]
for some constants $\alpha,\beta\ge0$. Moreover, if we also have $\tilde{H}(\mathrm{Bern}(1/2))=1$ and $\lim_{\rho\to0}\tilde{H}(\mathrm{Bern}(\rho))=0$, then $\tilde{H}(P)=H(P)$.
\end{thm}
\smallskip{}

\begin{IEEEproof}
Note that $\tilde{H}(P)\ge\tilde{H}(\emptyset)$ (where $\emptyset$ denotes a constant random variable) by monotonicity, and $\tilde{H}(\emptyset)=2\tilde{H}(\emptyset)$ by additivity, so $\tilde{H}(P)\ge\tilde{H}(\emptyset)=0$. We first prove
\begin{equation}
\tilde{H}(P)\le\gamma(P\Vert Q)\tilde{H}(Q).\label{eq:h_tilde_P_Q}
\end{equation}
To prove this, consider $\epsilon>0$. By the definition of $\gamma:=\gamma(P\Vert Q)$, there exists $n_{0}$ such that for every $n\ge n_{0}$, $m\ge(\gamma+\epsilon)n$, there exists $\bar{Q}\in\Gamma_{m}(Q)$ with $\bar{Q}\sqsubseteq P^{\otimes n}$. Take $m$ to be a power of $2$ satisfying $m\ge(\gamma+\epsilon)n_{0}$, and $n=\lfloor m/(\gamma+\epsilon)\rfloor$. We have $m\tilde{H}(Q)\ge\tilde{H}(\bar{Q})$, which holds either by ID-subadditivity, or by repeated applications of quasi-subadditivity since $m$ is a power of $2$. Hence, $m\tilde{H}(Q)\ge\tilde{H}(\bar{Q})\ge n\tilde{H}(P)$ by additivity and monotonicity. Taking $m\to\infty$ gives $(\gamma+\epsilon)\tilde{H}(Q)\ge\tilde{H}(P)$. The claim (\ref{eq:h_tilde_P_Q}) follows from $\epsilon\to0$.

Let $\beta:=\lim_{\rho\to0}\tilde{H}(\mathrm{Bern}(\rho))$ (which exists since $\tilde{H}(\mathrm{Bern}(\rho_{1}))\le\tilde{H}(\mathrm{Bern}(\rho_{2}))$ for $0<\rho_{1}\le\rho_{2}\le1/2$ by (\ref{eq:h_tilde_P_Q})), and $\alpha:=\tilde{H}(\mathrm{Bern}(1/2))-\beta$. Let 
\[
Q_{a,b,\rho}:=\mathrm{Bern}(1/2)^{\otimes a}\otimes\mathrm{Bern}(\rho)^{\otimes b}.
\]
We have $\lim_{\rho\to0}\tilde{H}(Q_{a,b,\rho})=a(\alpha+\beta)+b\beta$ by additivity. Consider distribution $P$. Take $a=\lfloor nH(P)\rfloor$, $b=\lfloor nH_{0}(P)\rfloor-a$. By (\ref{eq:h_tilde_P_Q}) and Theorem \ref{thm:gamma},
\begin{align*}
\tilde{H}(P) & \le\gamma(P\Vert Q_{a,b,\rho})\tilde{H}(Q_{a,b,\rho})\\
 & =\max\Big\{\frac{H(P)}{a+bH_{\mathrm{b}}(\rho)},\,\frac{H_{0}(P)}{a+b}\Big\}\tilde{H}(Q_{a,b,\rho})\\
 & \stackrel{(a)}{\to}\max\Big\{\frac{H(P)}{a},\,\frac{H_{0}(P)}{a+b}\Big\}(a(\alpha+\beta)+b\beta)\\
 & =\max\Big\{\frac{H(P)}{\lfloor nH(P)\rfloor},\,\frac{H_{0}(P)}{\lfloor nH_{0}(P)\rfloor}\Big\}(\alpha\lfloor nH(P)\rfloor+\beta\lfloor nH_{0}(P)\rfloor)\\
 & \stackrel{(b)}{\to}\alpha H(P)+\beta H_{0}(P),
\end{align*}
where we take $\rho\to0$ in (a), and we take $n\to\infty$ in (b). Also by (\ref{eq:h_tilde_P_Q}),
\begin{align*}
\tilde{H}(P) & \ge\frac{1}{\gamma(Q_{a,b,\rho}\Vert P)}\tilde{H}(Q_{a,b,\rho})\\
 & =\min\Big\{\frac{H(P)}{a+bH_{\mathrm{b}}(\rho)},\,\frac{H_{0}(P)}{a+b}\Big\}\tilde{H}(Q_{a,b,\rho})\\
 & \stackrel{(\rho\to0)}{\to}\min\Big\{\frac{H(P)}{a},\,\frac{H_{0}(P)}{a+b}\Big\}(a(\alpha+\beta)+b\beta)\\
 & \stackrel{(n\to\infty)}{\to}\alpha H(P)+\beta H_{0}(P).
\end{align*}
Hence, $\tilde{H}(P)=\alpha H(P)+\beta H_{0}(P)$.
\end{IEEEproof}
\smallskip{}

\section{Acknowledgement}

The work of C.\ T.\ Li was supported in part by a grant from the Research Grants Council of the Hong Kong Special Administrative Region, China [Project No: CUHK 14209823 (GRF)]. The work of S.\ Asoodeh was partially supported by the Natural Sciences and Engineering Research Council of Canada.

\smallskip{}

\appendix{}

\subsection{Proof of Proposition \ref{prop:properties}}\label{subsec:pf_properties}

For the lower bound (\ref{eq:prop_lower_bound}), consider $X,Y_{1},\ldots,Y_{m}$ with $X\sim P$, $Y_{i}\sim Q$ and $H(X|Y^{m})=0$. We have $H(P)=H(X)\le H(Y^{m})\le mH(Q)$, and also $H_{0}(P)\le mH_{0}(Q)$ similarly. To show (\ref{eq:prop_lower_bound_infty}), consider $\bar{Q}\in\Gamma_{m}(Q)$ with $\bar{Q}\sqsubseteq P$, so $\max_{y^{m}}\bar{Q}(y^{m})\le\max_{x}P(x)$. Assume $Q(0)=\max_{y}Q(y)$. Union bound gives $1-\bar{Q}(0,\ldots,0)\le m(1-Q(0))$, so 
\begin{align*}
m & \ge\frac{1-\bar{Q}(0,\ldots,0)}{1-Q(0)}\\
 & \ge\frac{1-\max_{y^{m}}\bar{Q}(y^{m})}{1-Q(0)}\;\ge\;\frac{1-\max_{x}P(x)}{1-Q(0)}.
\end{align*}

For the chain rule, let $m:=m^{*}(P\Vert R)$, $m':=m^{*}(R\Vert Q)$. Consider $P_{Z^{m}|X}$ which achieves $m^{*}(P\Vert R)$ (i.e., if $X\sim P$, $Z^{m}|X\sim P_{Z^{m}|X}$, then $Z_{i}\sim R$ and $H(X|Z^{m})=0$), and $P_{Y^{m'}|Z}$ which achieves $m^{*}(R\Vert Q)$. Generate $X\sim P$, $Z^{m}|X\sim P_{Z^{m}|X}$ and $(Y_{i,1},\ldots,Y_{i,m'})|Z_{i}\sim P_{Y^{m'}|Z}$ for $i\in[m]$. We have $H(X|Z^{m})=H(Z^{m}|(Y_{i,j})_{i,j})=0$, so $H(X|(Y_{i,j})_{i,j})=0$.

For subadditivity, let $m:=m^{*}(P_{X}\Vert Q)$, $m':=m^{*}(P_{Z}\Vert Q)$. Consider $P_{Y^{m}|X}$ which achieves $m^{*}(P_{X}\Vert Q)$, and $P_{Y^{m'}|Z}$ which achieves $m^{*}(P_{Z}\Vert Q)$. Generate $(X,Z)\sim P_{X,Z}$, $Y^{m}|X\sim P_{Y^{m}|X}$, $\tilde{Y}^{m'}|Z\sim P_{Y^{m'}|Z}$. We have $H(X|Y^{m})=H(Z|\tilde{Y}^{m'})=0$, so $H(X,Z|Y^{m},\tilde{Y}^{m'})=0$.

We now prove almost quasiconvexity. Let $m:=m^{*}(P_{Z}\Vert P_{Y})$, $n:=\max_{z}m^{*}(P_{X|Z=z}\Vert P_{Y|Z=z})$. By the definition of $m^{*}(P_{Z}\Vert P_{Y})$, we can consider a coupling $Y_{1},\ldots,Y_{m}$ in $\Gamma_{m}(P_{Y})$ (i.e., $Y_{i}\sim P_{Y}$) and a function $\tilde{Z}:=f(Y_{1},\ldots,Y_{m})$ such that $\tilde{Z}\sim P_{Z}$. Conditional on the event $\tilde{Z}=z$, generate $\tilde{Y}_{1},\ldots,\tilde{Y}_{n}$ according to a coupling in $\Gamma_{n}(P_{Y|Z=z})$ (i.e., $\tilde{Y}_{i}|\{\tilde{Z}=z\}\sim P_{Y|Z=z}$) where there is a function $\tilde{X}:=g_{z}(\tilde{Y}_{1},\ldots,\tilde{Y}_{n})$ with $\tilde{X}|\{\tilde{Z}=z\}\sim P_{X|Z=z}$. Since $\tilde{Z}\sim P_{Z}$, we have $\tilde{X}\sim P_{X}$. Also, since $\tilde{Y}_{i}|\{\tilde{Z}=z\}\sim P_{Y|Z=z}$ and $\tilde{Z}\sim P_{Z}$, we have $\tilde{Y}_{i}\sim P_{Y}$. Hence, $Y_{1},\ldots,Y_{m},\tilde{Y}_{1},\ldots,\tilde{Y}_{n}$ is a coupling in $\Gamma_{m+n}(P_{Y})$. The result follows from $\tilde{X}=g_{f(Y_{1},\ldots,Y_{m})}(\tilde{Y}_{1},\ldots,\tilde{Y}_{n})$ being a function of $(Y_{1},\ldots,Y_{m},\tilde{Y}_{1},\ldots,\tilde{Y}_{n})$.

\smallskip{}

\subsection{Proof of Lemma \ref{lem:alias}}\label{subsec:pf_alias}

\cite[Lemma 1]{li2021efficient} shows that for probability vectors $p,r\in\mathbb{R}^{n}$ with entries in descending order and $r\preceq p$, there is a lower triangular right stochastic matrix $M\in\mathbb{R}^{n\times n}$ where each row has at most one positive off-diagonal entry, and $p=rM$. We show that this implies the statement of Lemma \ref{lem:alias} for completeness. Assume $P,R$ are supported within $\mathbb{N}$ with probabilities in descending order. We first consider the case where $P,R$ are finite and supported within $[n]$, so we can define the corresponding probability vectors $p,r$. Generate $Z\sim R$, generate $X$ conditional on $Z$ according to the conditional probability matrix $M$, and let $W=\mathbf{1}\{X=Z\}$, with $H(X|Z,W)=0$ since there is at most one off-diagonal entry (the conditional distribution of $X$ given $Z=z$ is at most binary). We have $W\sim\mathrm{Bern}(\rho)$ for some $\rho\in[0,1]$. If $\rho\le1/2$ is desired, we can simply consider either $W$ or $1-W$.

This also applies when $R$ has infinite support as long as $P$ is finite (assume its support is $[n]$). To show this, fix $n'\in\mathbb{N}$, generate $Z\sim R$, and let $\hat{Z}:=\min\{Z,n'\}$. Let $\hat{R}$ be the distribution of $\hat{Z}$. We have $\sum_{i=1}^{k}\hat{R}^{\downarrow}(i)=\sum_{i=1}^{k}R(i)$ for $k\in[n]$ as long as $\sum_{i=n'}^{\infty}R(i)\le R(n)$, which holds for large enough $n'$. Since $R\preceq P$, we also have $\hat{R}\preceq P$, so we can apply the lemma on $\hat{R}$ instead.

\smallskip{}

\bibliographystyle{IEEEtran}
\bibliography{ref}

\end{document}